\documentclass[twocolumn,tighten]{aastex62}

\usepackage{blindtext}
\usepackage{graphicx}
\usepackage[colorinlistoftodos]{todonotes}
\usepackage{lineno}

\begin{document}

\title{AGN jets and a fanciful trio of black holes in the Abell 85 Brightest Cluster Galaxy}

\author{Juan P.\ Madrid}

\affiliation{Department of Physics and Astronomy, The University of Texas Rio Grande Valley, Brownsville, TX 78520, USA}

                         
\begin{abstract}

A new radio map of the Abell 85 Brightest Cluster Galaxy (BCG) was obtained with the Karl G. 
Jansky Very Large Array (VLA). With a resolution of 0.$\arcsec$02, this radio image shows 
two  kiloparsec-scale bipolar jets emanating from the active galactic nucleus of the Abell 85 BCG. 
The galaxy core appears as a single entity on the new radio map. It has been assumed that the Abell 85 BCG 
contained a binary black hole in its core. However, Chandra X-ray data and the new high-resolution 
radio map show no evidence that the Abell 85 BCG harbors a binary black hole. The assumption 
that this galaxy contains a binary black hole was based on the analysis of its optical surface 
brightness profile obtained under poor seeing conditions. We demonstrate how the well-known 
blurring effects of atmospheric seeing can mimic the effects of a binary supermassive black 
hole (SMBH). Likewise, SDSS J004150.75-091824.3 was postulated to be ``a third" SMBH associated 
with the BCG. In the optical and X-rays, \object{SDSS J004150.75-091824.3} is a  point-like 
source located $\sim14\arcsec$ away from the nucleus of the Abell 85  BCG. A new spectrum of 
SDSS J004150.75-091824.3, obtained with the 10.4-m Gran Telescopio Canarias, reveals that 
this source is a background quasar at a redshift of $z=1.560\pm 0.003$ and  not associated 
in any way with the Abell 85 cluster. 

\end{abstract}



\section{Introduction}
\subsection{The Abell 85 Galaxy Cluster}

Abell 85 is a galaxy cluster that is in the process of merging with two 
subclusters \citep{ichinohe2015}. As a bright X-ray cluster with a complex structure, 
Abell~85 has been intensively studied with all modern X-ray observatories: Chandra, 
XMM-Newton and Suzaku \citep[e.g.][]{kempner2002, durret2003, schenck2014}. 

The galaxy population of Abell 85 was recently analyzed at optical wavelengths by \citet{agulli2016} 
who determined the spectroscopic luminosity function with 460 confirmed cluster members. 

\citet{owen1997} and \citet{schenck2014} made clear detections of radio emission emanating from
the Abell 85 cluster and its BCG. Radio relics located $\sim$320 kpc from the core of Abell 85 
were discovered by \citet{slee2001} -- see also \citet{schenck2014} and \citet{bagchi1998}. 
These relics are likely the result from shocks of past cluster mergers.
\subsection{A fanciful trio of black holes}
\begin{figure*}
        \centering
        \begin{tabular}{cc} 
        \includegraphics[scale=0.45]{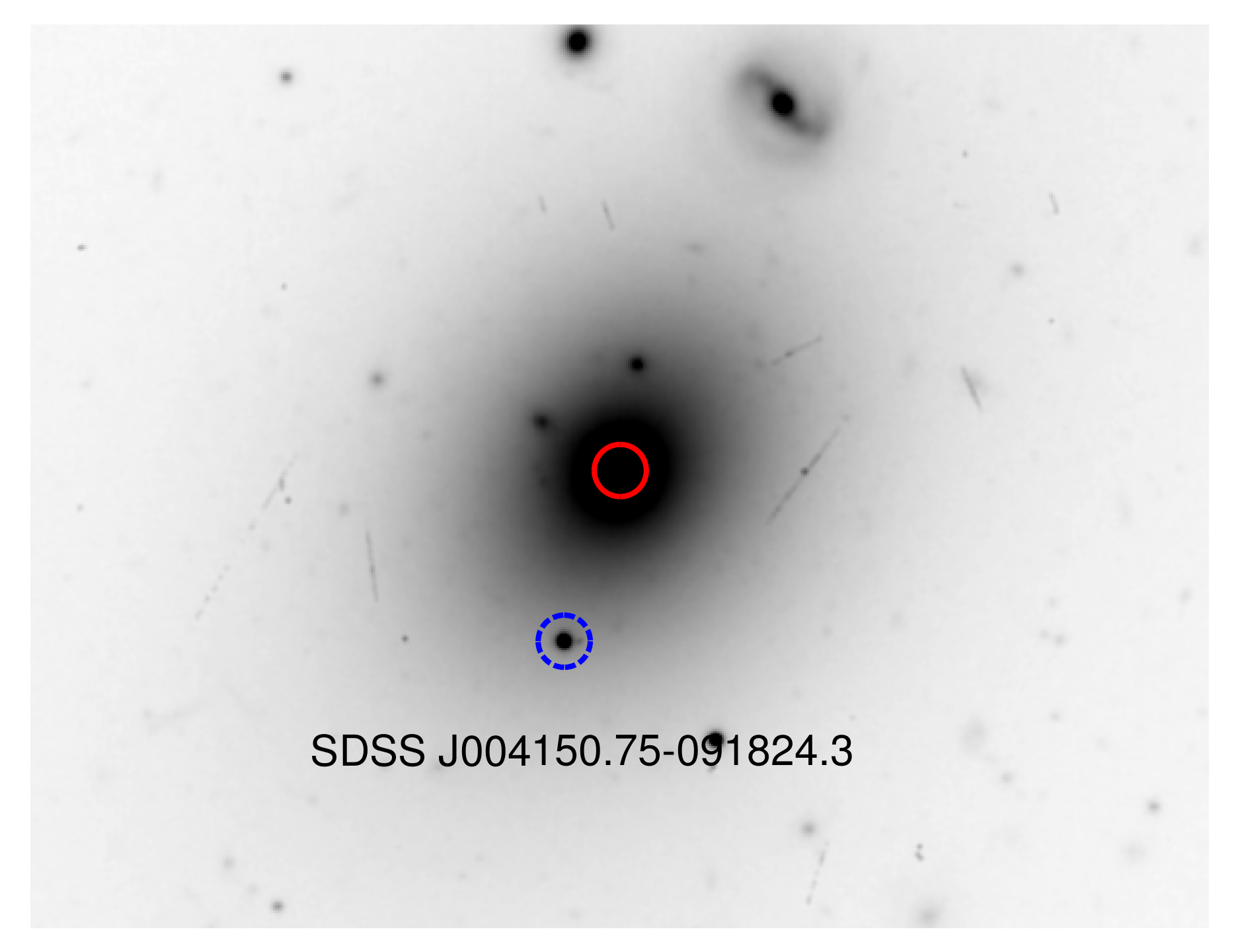} &  \includegraphics[scale=0.45]{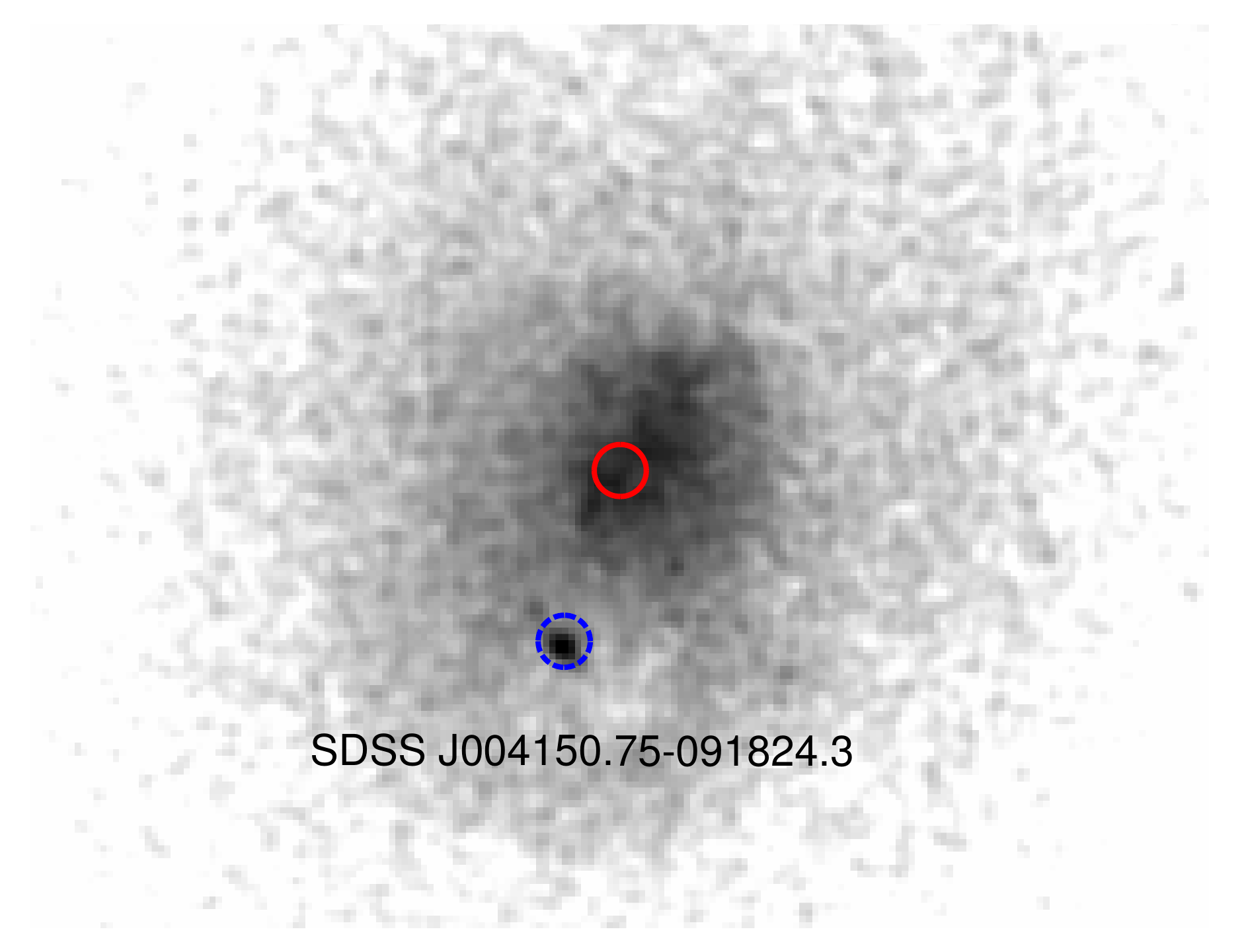}\\
        \end{tabular}
   \caption{Central region of Abell 85 as viewed by Gemini (left) and Chandra (right). The center of the Abell 85 BCG
    is depicted with a red solid circle.  The location of SDSS J004150.75-091824.3 is shown with a blue dotted circle.
    These circles have diameters of 4 arcseconds. The optical image was originally presented in  \citet{madrid2016} 
    while the X-ray data was presented in \citet{ichinohe2015}. North is up and east is left. X-ray data kindly provided 
    by Y. Ichinohe.\label{geminichandra}}
\end{figure*}

In the optical and X-rays, SDSS J004150.75-091824.3 is a bright ($g$=14.03 mag), point-like 
source located $\sim14\arcsec$ away from the nucleus of the Abell 85  BCG -- see 
Fig.~\ref{geminichandra}. SDSS J004150.75-091824.3 is also the closest X-ray point source 
to the nucleus of the BCG. The core of the BCG shows a diffuse X-ray emission while 
SDSS~J004150.75-091824.3 is bright and point-like.

SDSS J004150.75-091824.3 was included in the photometric selection of quasars from the SDSS 
by \citet{richards2004,richards2009}. SDSS J004150.75-091824.3 was found to have a photometric 
redshift of $z$= 0.925 in \citet{richards2004}. This value was later revised to $z$=0.675 
in \citet{richards2009}. 

SDSS J004150.75-091824.3 was identified as a possible AGN in the detailed X-ray study carried out by 
\citet{durret2005}. These authors identify SDSS J004150.75-091824.3 as an X-ray point source on their 
Chandra data and also identify an optical counterpart on a SDSS image (see their Fig.\ 2).
SDSS J004150.75-091824.3 is located within an X-ray cavity \citep{durret2005, ichinohe2015}.

Given the projected proximity of SDSS J004150.75-091824.3 to the center of the Abell 85 BCG, 
\citet{lopez2014} surmised that stellar light could have biased the redshift and classification 
given by \citet{richards2009}. \citet{lopez2014} speculated that \textsc{SDSS J004150.75-091824.3} 
could be ``a third" supermassive black hole associated with the Abell 85 brightest cluster galaxy. 
The other two black holes of Abell 85 would have been a hypothetical binary pair in the 
core of the BCG.  \citet{lopez2014} claimed the presence of a substantial light deficit on the nuclear 
region of the  Abell 85 BCG. \citet{lopez2014} interpreted this light deficit  as evidence 
for the presence of an ``ultramassive" SMBH. 

When galaxies merge, their central black holes are  thought to consolidate into one entity. 
During the process of black hole coalescence, supermassive black hole binaries are thus created 
\citep{begelman1980}.  Stars that come in close proximity to a binary supermassive black hole 
(SMBH) can be slingshot away through gravitational interactions \citep{begelman1980, ebisuzaki1991, milo2001}. 
Over time, as this process repeats itself, a SMBH creates a deficit of stars in its vicinity. 

The presence of a binary SMBH can then be inferred by the detection of stellar light deficits on 
the optical surface brightness profiles of, for instance, elliptical galaxies 
\citep[e.g.][]{graham2004,dullo2019}. When light deficits are present in surface brightness 
profiles, the steep slope of the profile in the outer parts of the galaxy becomes flatter, 
sometimes even close to constant, with decreasing radii \citep[e.g.][]{graham2004}.
 
Given the small spatial scales involved, the accurate characterization of the properties
of stellar cores and, among them, their light deficits, require Hubble Space Telescope data
\citep{ferrarese2006}. Light deficits associated with SMBH occur in spatial scales of a few 
hundred parsecs in galaxies located megaparsecs away. For instance, cores in galaxies belonging 
to the Virgo cluster have spatial scales of 1 to 4 arcseconds \citep{ferrarese1994}. 

With few exceptions  \citep[e.g., 3C~75;][]{owen1985}, unequivocal detections of close pairs 
of binary AGN remain rare. Binary black holes are elusive, even when dedicated observing 
campaigns are carried out \citep{tingay2011}.

In the following sections a new radio map of the Abell 85 BCG with the highest resolution 
ever achieved is introduced. This radio map is used to search for the presence of a binary AGN.
A new optical spectrum of SDSS J004150.75-091824.3 is also presented. This spectrum is  used 
to make a precise determination of the redshift for this source. A new Gemini spectrum of 
the Abell 85 BCG is also shown. High-resolution imaging obtained with Gemini and Chandra 
are used as ancillary data to place the new observations in a wider context. Finally, by 
presenting optical data obtained under different atmospheric conditions we illustrate the 
effect of seeing on optical surface brightness profiles.

At the distance of Abell 85 ($D=233.9$ Mpc) 1$\arcsec$ corresponds to 1.075 kpc.

\section{Observations and data reduction}
\subsection{X-ray Imaging}
The Chandra data shown in Fig.~\ref{geminichandra}  was analyzed and presented by \citet{ichinohe2015}. 
Details of the observations and data reduction for this dataset are given in that reference.
 
\citet{ichinohe2015} created a Chandra image that is a factor of $\sim$5 deeper than the initial Chandra 
image of $\sim$37 ks analyzed by \citet{kempner2002}. \citet{ichinohe2015} also included XMM–-Newton and 
Suzaku data out to the virial radius of the cluster. \citet{ichinohe2015} found evidence that the 
Abell 85 cluster is undergoing two mergers and has  gas sloshing out to a radius of $r\sim600$ kpc from 
its center.
\subsection{Optical Imaging}
Figure \ref{geminichandra} also shows a Gemini  image of the Abell 85 BCG. The Gemini $r$-band
image was obtained with the Gemini Multi-Object Spectrograph (GMOS). This dataset was presented 
and discussed in depth in \citet{madrid2016}. A new Gemini image obtained under poor seeing conditions
is also presented in section 4  (Gemini program  GS-2016B-Q-87). The data reduction for this new 
image is identical to the one described in \citet{madrid2016}.

\begin{figure}
 \epsscale{1.3}
 \plotone{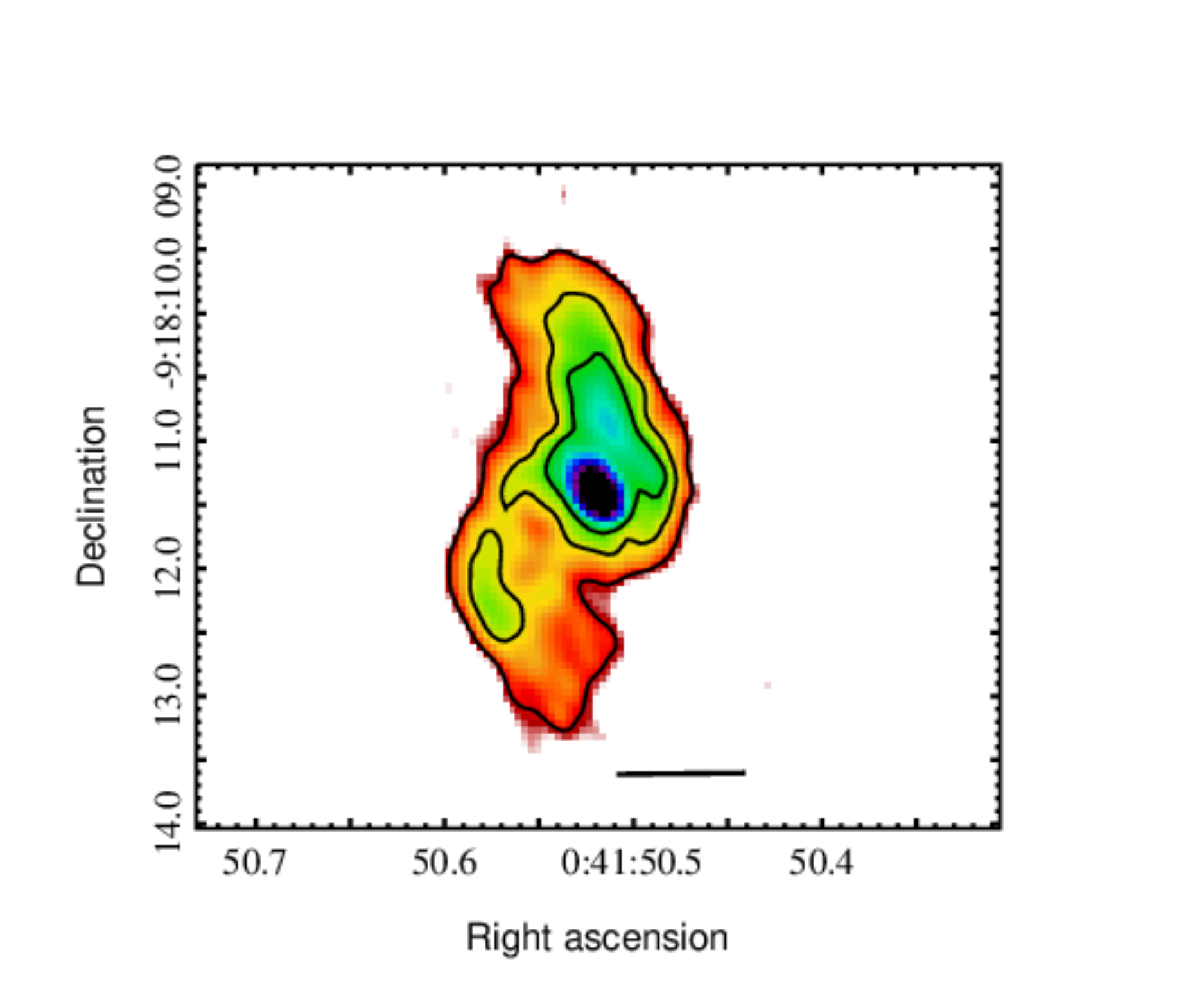}
 \caption{VLA synthesized image of the Abell 85 BCG. This image shows two bipolar jets emanating from the core
 of Abell 85. Contours are shown at 0.02, 0.06, and 0.1 mJy. The scale bar on the lower right is 1$\arcsec$ in length. 
 North is up and east is left.
 \label{vlamap}}
\end{figure}
\begin{figure*}
 \epsscale{0.85}
 \plotone{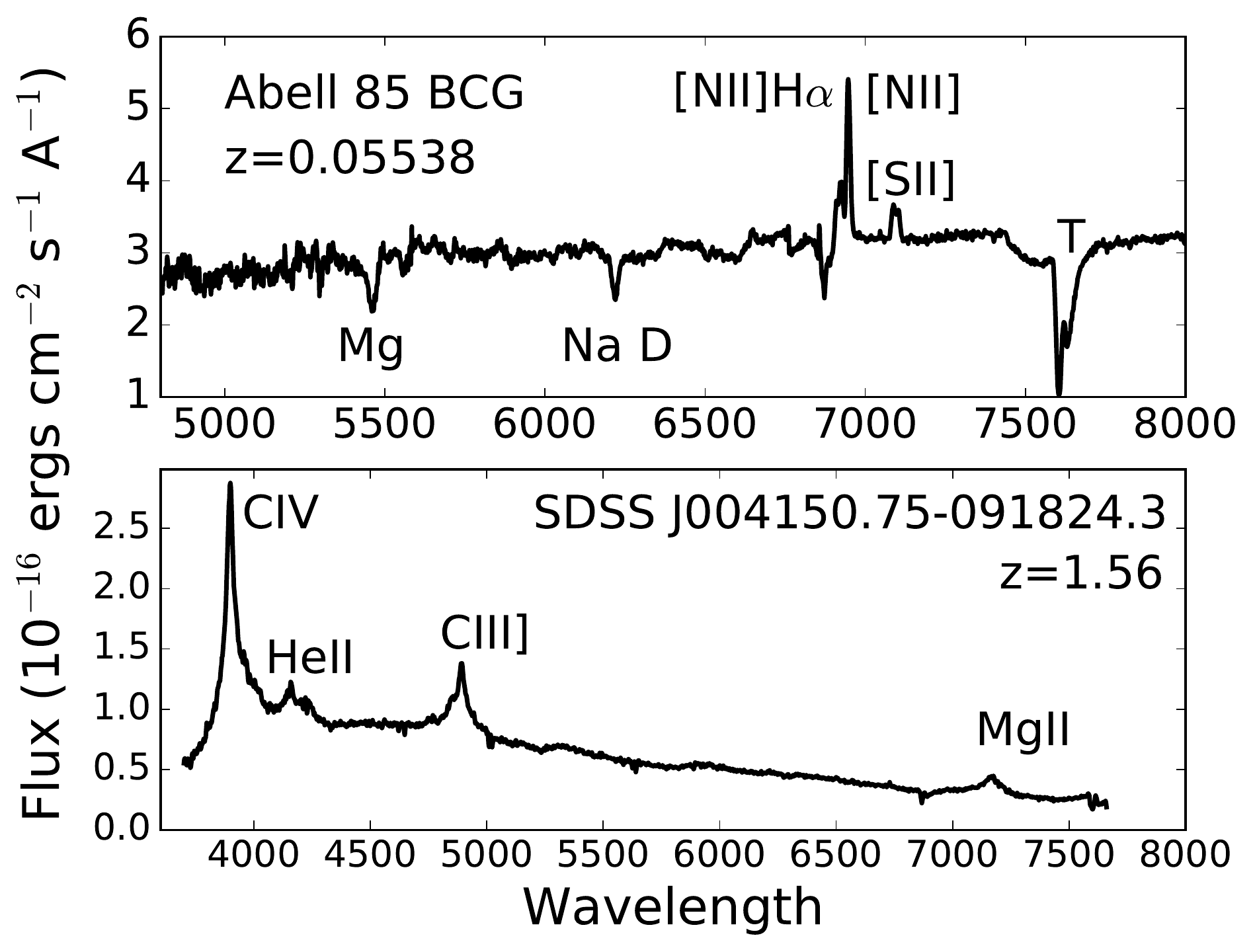}
 \caption{\textit{Top:} Gemini spectrum of the Abell 85 BCG. \textit{Bottom:} GTC spectrum of SDSS J004150.75-091824.3. 
 Both spectra are shown on their observed wavelength without correcting for radial velocities. The letter \textsc{T} 
 denotes a telluric absorption line.
 \label{spectra}}
\end{figure*}
\subsection{Optical Spectroscopy}
The Gemini spectrum of the Abell 85 BCG was obtained  with GMOS-South on spectroscopic mode. 
Five exposures of 900 seconds were obtained on 2016 September 14. The grating used was R400 which
has its blaze wavelength at 764 nanometers. The slit width was 1.$\arcsec$5.

The first task of the data reduction process is to apply {\sc gprepare} to the raw science data 
obtained through the Gemini archive. {\sc gprepare} applies a flatfield and performs overscan 
subtraction. The spectrum is extracted from the fits file by using the task {\sc gsextract}. 
This tasks extracts the intensity values along the spectral dimension of the CCD chip. 
Calibration is carried out with the task {\sc gswavelength}. The details of the reduction procedure 
used here were described in \citet{madrid2013}.\\

The spectroscopic data for SDSS J004150.75-091824.3 was obtained with the Optical System for Imaging and 
low Intermediate Resolution Integrated Spectroscopy (OSIRIS) on the Gran Telescopio Canarias. OSIRIS
is an optical imager and spectrograph. A $2\times2$ binning was used for the CCD resulting in a spatial 
scale of 0.$\arcsec$26 per pixel. Likewise, the resulting spectral resolution is 2.12~\AA/pixel. 
The slit had a width of 1$\arcsec$ and  the grating in use was R1000B.

The spectrum of SDSS J004150.75-091824.3 presented here was obtained on 2017 August 17. Raw science data,  
bias, flats and arcs were retrieved from the Grantecan public archive.

Both science data extraction and sky subtraction were done using the \textsc{pyraf} task \textsc{apall}. 
In the same way, arcs, bias and flats are processed with \textsc{apall}. The science spectra had the bias 
removed and were normalized by the flat. The HgAr and Ne calibration arcs were used for wavelength 
calibration. The emission lines given on the reference arcs provided by the observatory were used 
as part of the wavelength calibration. The spectra were extracted using a spatial width of 7 pixels, 
or 1.$\arcsec$8. The spectrophotometric standard star {\sc Feige 110} is used to flux calibrate the spectra
using the tasks \textsc{standard, sensfunction} and \textsc{calibrate}. 
\subsection{Radio Data}
Karl G. Jansky Very Large Array (VLA) data was obtained from the National Radio Astronomy 
Observatory public archive. Abell 85 was observed with the VLA in its A configuration on 
2018 March 25. The total observation time was 4 hours, resulting in 2.8 hours on-source 
after overheads. The observations used the X-band receiver with 4 GHz of correlated 
bandwidth centered at 10 GHz.

The VLA data were processed with a series of tasks found within the Common Astronomy 
Software Applications (CASA). Standard interferometric data reduction procedures were 
followed to calibrate the delay, bandpass, phase, amplitude and flux. One round of 
phase-only self-calibration was also included.   During imaging, a Briggs robust 
weighting of $+0.5$ was used to suppress Point Spread Function (PSF) sidelobes and 
achieve a resolution of $0.\arcsec29 \times 0.\arcsec19$ at a position angle 
of 32 degrees. The final image had a measured sensitivity of 3 $\mu$Jy/beam.

\section{The Abell 85 AGN}

The high-resolution and sensitivity of the radio data allows the detection of the central AGN 
and bipolar diffuse emission. Compact radio sources are ubiquitous in the core of early type 
elliptical galaxies \citep{ekers1973}, and the Abell 85 BCG is no exception. The VLA image 
also shows two bipolar jets emanating from the compact core. Both jets are less than $\sim$2 kpc 
in length (projected). The image shown in Fig.~\ref{vlamap} is the first radio image
to clearly identify the jets emanating from the Abell 85 BCG.

Interestingly, the southern jet is well aligned with the location of the X-ray cavity 
described by \citet{durret2005} and  \citet{ichinohe2015}; see Fig.~\ref{geminichandra}. 
This X-ray cavity is, however, at about 20 kpc south of the core whereas the southern 
radio jet does not show radio emission beyond 2 kpc from the center of the BCG.

The VLA image of the Abell 85 BCG shows a single core and no evidence of double nuclei. 
Similarly, the X-ray image only shows diffuse X-ray emission in the core of the galaxy. 
Binary black holes, when present, appear as two distinct point sources in X-ray images 
\citep{komossa2003, fabbiano2011}. 

The null detection of binary black holes in the X-ray and very high resolution radio data 
implies that the central massive object in the Abell 85 BCG is likely a single 
entity. Should a binary AGN exist in the  Abell 85 BCG, its detection would require very 
long baseline interferometry (VLBI) given that the pair would be separated by less than 
$\sim$400 pc. Using data obtained with the Very Long Baseline Array (VLBA), \citet{rodriguez2006} 
discovered a supermassive binary black hole in the radio galaxy IVS 0402$+$379 
(B1950)\footnote{IVS stands for International VLBI Service. The VLBA Calibrator Survey
gives the (J2000.0) name for this source: J0405+3803 \citep{beasley2002}.},
the projected separation between the two black holes is only 7.3 pc.\\

\section{SDSS J004150.75-091824.3 is a distant quasar}

Three star-like sources are present in the optical image within 14$\arcsec$ (or 15.05 kpc) 
of the nucleus of the BCG. One of these sources is SDSS J004150.75-091824.3, see 
Fig.~\ref{geminichandra}. On the other hand, in the X-ray image, SDSS J004150.75-091824.3 
is the only point-like source within one arcminute of the center of Abell 85. With the 
Gemini image we can derive an accurate position for SDSS J004150.75-091824.3 at R.A.=0:41:50.764 
and dec=$-$9:18:24.4.

The core of Abell 85 has a diffuse and extended X-ray emission that envelops SDSS~J004150.75-091824.3. 
The presence of this extended X-ray emission was misinterpreted by \citet{lopez2014} as an 
indication that SDSS J004150.75-091824.3 was a low redshift AGN.

In Fig.~\ref{spectra}, the spectrum of SDSS J004150.75-091824.3 show four prominent emission 
lines: C\,{\sc iv}, He\,{\sc ii}, C\,{\sc iii}], and Mg\,{\sc ii}. These emission lines are 
characteristic of quasar spectra, e.g.,\citet{wilkes1986, vandenberk2001}. With the Grantecan 
spectrum  the redshift of SDSS J004150.75-091824.3 is determined to be $z=1.5603\pm 0.003$.

The spectrum of the Abell  85 BCG, also shown in Fig.~\ref{spectra},  has a prominent 
H$\alpha~+$ [N\,{\sc ii}] 6584  emission line. The H$\alpha$ line has a [N\,{\sc ii}] 
emission line on either side. The forbidden [S\,{\sc ii}]  6717/6730 doublet is also 
clearly visible. The Franhofer absorption lines for magnesium and sodium are  easily seen. 
The above emission and absorption lines are used to confirm the redshift of the BCG to 
be $z=0.05538 \pm 0.0004$. This value is in agreement with several earlier determinations 
of the redshift of the  Abell  85 BCG \citep[][among others]{nesci1990}.

It should be noted that the spectrum of the BCG is similar to the archival spectrum of this 
galaxy available through the SDSS archive. The spectrum of the Abell 85 BCG is presented 
here as a comparison with the spectrum of SDSS J004150.75-091824.3 but also as a  baseline 
spectrum for studies of this galaxy, e.g.,~\citet{edwards2016}.

\section{Mirage Supermassive Black Holes}
\subsection{The Effect of Seeing on Optical Surface Brightness Profiles}
The presence of a binary black hole in the Abell 85 BCG was postulated based on the 
analysis of its surface brightness profile \citep{lopez2014}. However, the obvious effect 
of poor seeing has been ignored in recent studies that attempt to use ground-based data 
without sufficient resolution to study the surface brightness profile of the Abell 85 BCG. 
This section highlights the risks of inferring the presence of a light deficit, or any 
other nuclear property of a distant galaxy, using ground-based data.

The BCG of Abell 85 has been reported during recent years as having both a
light deficit and a light excess in its core \citep{lopez2014, bonfini2015, madrid2016, mehrgan2019}. 
These apparently contradicting results are a clear example of the effects of data quality 
on optical surface brightness profiles.

\begin{figure}
        \centering
        \begin{tabular}{c}    
                \includegraphics[scale=0.4]{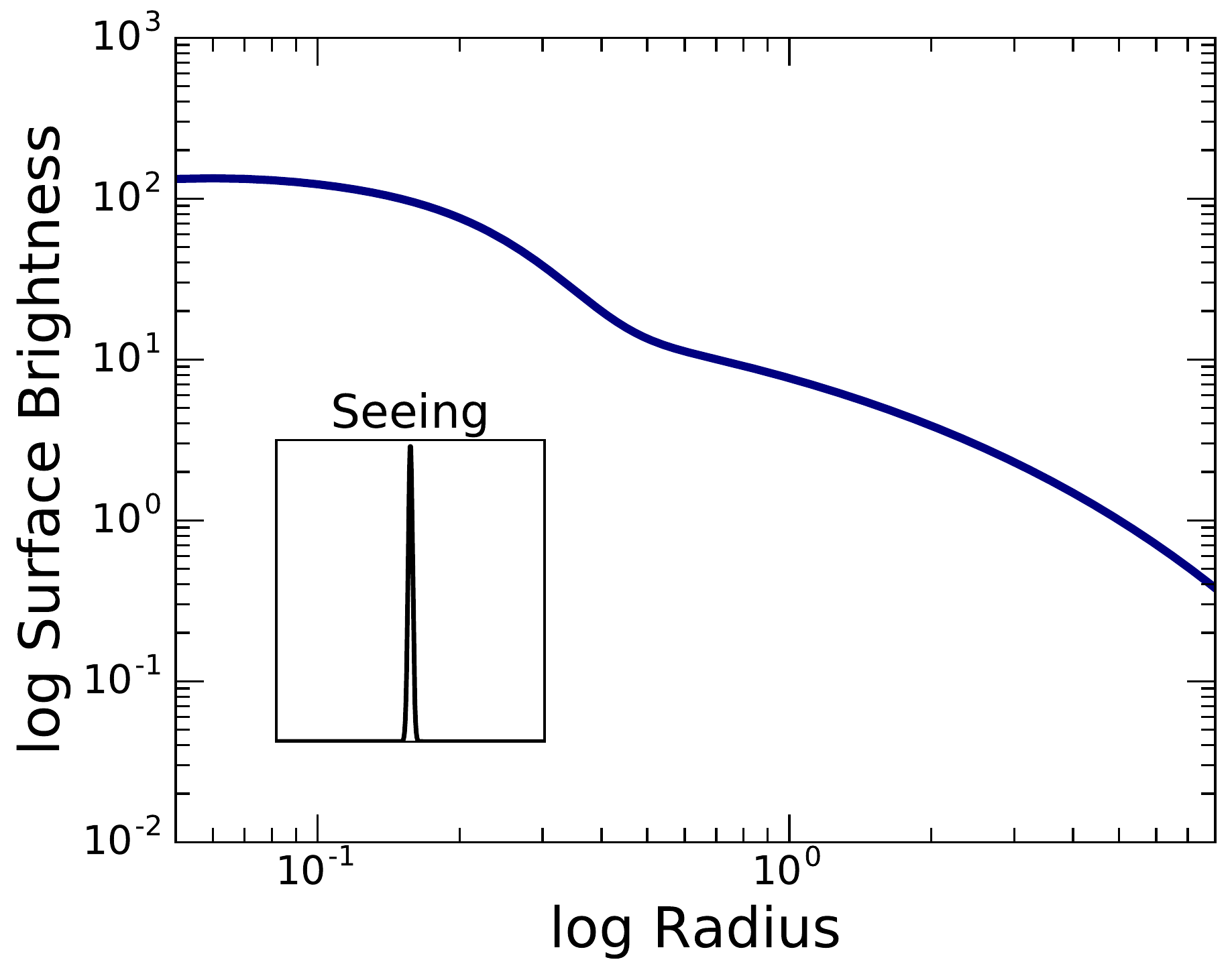}\\
                \includegraphics[scale=0.4]{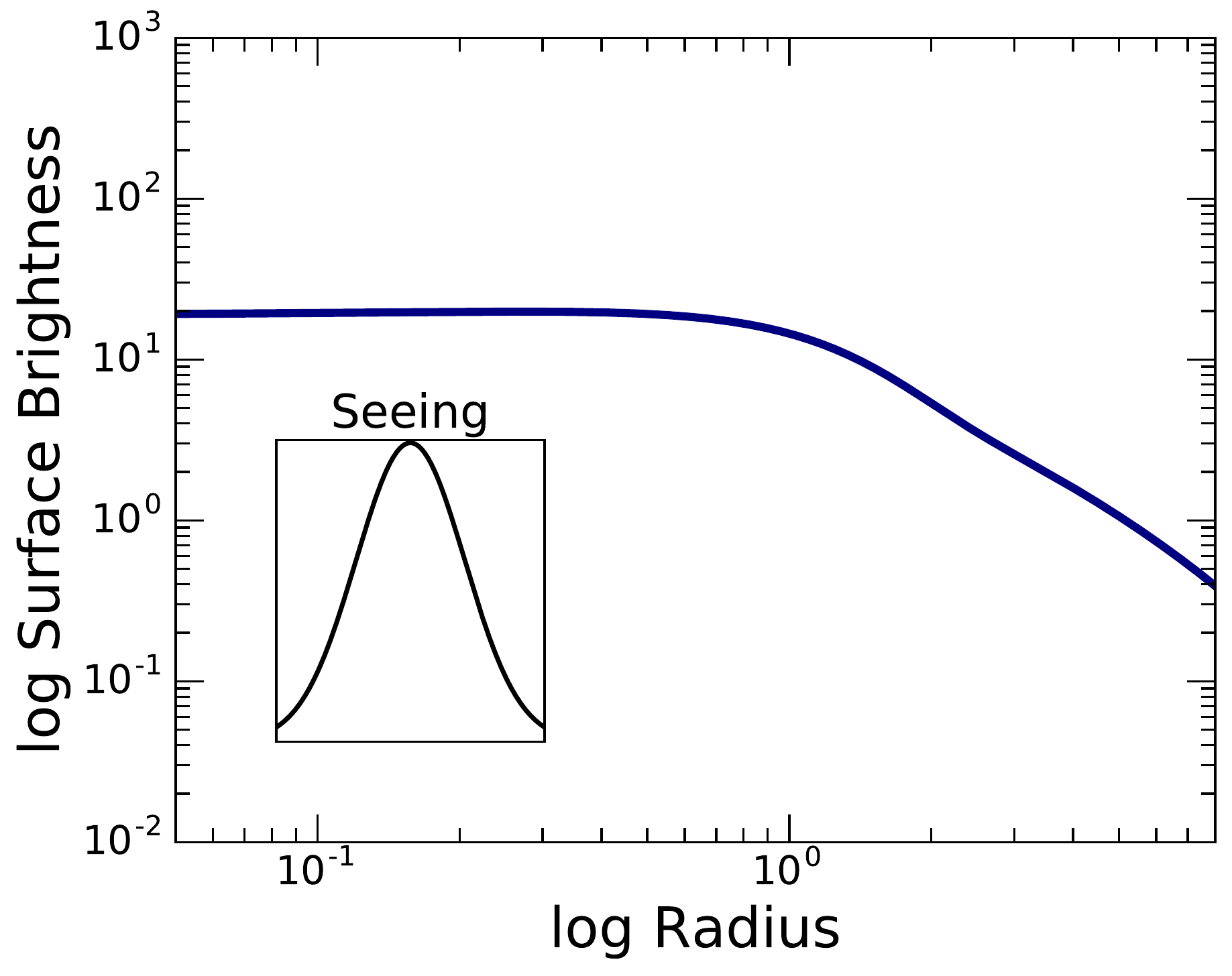} \\
        \end{tabular}
    \caption{Surface brightness profile of a model galaxy observed with optimal seeing (\textit{top panel}) 
    or with bad seeing (\textit{bottom panel}). The seeing is represented by a PSF that has been normalized. 
    Units are arbitrary. 
\label{seeingconvolve}
}
\end{figure}

The effect of seeing on surface brightness profiles can be easily modeled, as we do 
in this section. Atmospheric turbulence blurs and smears images obtained with ground 
based telescopes. Poor seeing degrades and erases the innermost structure of surface brightness 
profiles. This effect is not only intuitive but was well defined over four decades ago 
by \citet{schweizer1979}, among others. More recently, \citet{cote2006} showed
that data with poor seeing miss the presence of structures in galactic nuclei.

Let's consider a galaxy whose light profile can be described by two main components:
a nuclear point spread function and an outer spheroid. The nuclear component can be 
accurately modeled as a Gaussian and the main component can be represented by a 
S\'ersic function  \citep{sersic1968}. 

The surface brightness profile of the above galaxy observed with a ground based 
telescope can be evaluated by convolving its profile with the seeing. Here, seeing is 
considered as the distorting effects of both atmospheric turbulence and telescope optics. 

\begin{figure*}
\epsscale{0.7}
 \plotone{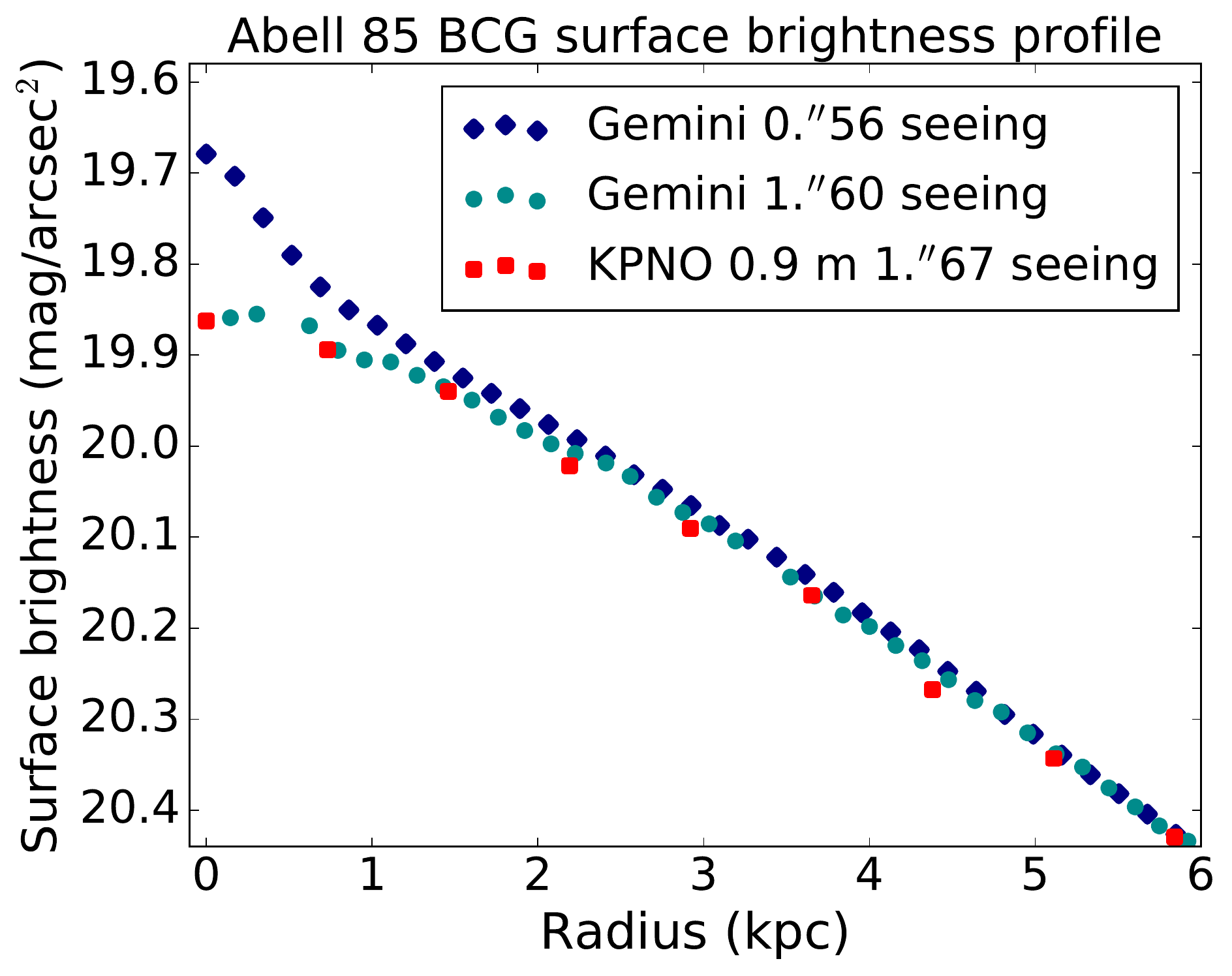}
 \caption{Surface brightness profile of the Abell 85 BCG obtained with Gemini and the KPNO 0.9-m.
 Two epochs of observations with Gemini are plotted. One epoch with excellent seeing (0.$\arcsec$56) 
 and a second epoch of observations obtained under poor weather conditions with a seeing of (1.$\arcsec$60). 
 \label{abell85sb}}
\end{figure*}


Fig.~\ref{seeingconvolve} shows the effect of seeing on the innermost regions of a surface 
brightness profile. Observing a galaxy light profile with excellent seeing recovers
the actual profile with no distortions, as shown in the top panel of Fig.~\ref{seeingconvolve}.
Both the nuclear and spheroidal components are detected when the light profile is obtained 
under ideal seeing conditions. 

On the other hand, when the same light profile is observed with poor seeing, its central 
part is entirely misrepresented, as shown in the bottom panel of Fig.~\ref{seeingconvolve}.
The central nuclear component is fully blurred by seeing effects. Under bad seeing conditions, 
instead of detecting a central Gaussian component the slope of the profile becomes 
flat toward the nucleus of the galaxy.

A surface brightness profile observed with bad seeing could give the illusion of a light deficit. 
Poor seeing also alters the light profile creating a break on its slope. The above can mislead 
observers to believe on the detection of a depleted core and its associated SMBH when galaxies 
are observed with poor seeing conditions.
\subsection{Optical observations of Abell 85}
\citet{lopez2014} used data obtained with the Kitt Peak National Observatory (KPNO) 0.9-meter 
telescope to analyze the central surface brightness profile of the Abell 85 BCG. The seeing 
for the KPNO 0.9-m data is reported to be 1.$\arcsec$67. \citet{bonfini2015} published 
a detailed analysis of the surface brightness profile of Abell 85 using data obtained with
the Canada-France-Hawaii telescope (CFHT). The CFHT data analyzed by  \citet{bonfini2015} has 
subarcsecond resolution: 0.$\arcsec$74. \citet{bonfini2015} found that the light deficit 
claimed by \citet{lopez2014} did not exist. Moreover, contrary to a light deficit, \citet{bonfini2015} 
find a modest light excess in the core of the Abell 85 BCG.

\citet{madrid2016} obtained Gemini data of the Abell 85 BCG with a seeing of 0.$\arcsec$56. 
\citet{madrid2016} confirmed the presence of a light excess in the core of this galaxy initially
found by \citet{bonfini2015}. The superior seeing of the Gemini data allow \citet{madrid2016} 
to detect an additional nuclear component within the inner 2 kpc of the center of the Abell 85 BCG. 
This nuclear component is well resolved, with a FWHM of 0.$\arcsec$85. At the distance of Abell 85
this nuclear component has a size of $\sim$0.9 kpc \citep{madrid2016}.

The empirical effect of seeing on the central surface brightness profile of the Abell 85 BCG is shown 
in Fig.~\ref{abell85sb}. This figure shows the light profile derived with Gemini data obtained 
during two different epochs with different seeing conditions. Fig.~\ref{abell85sb} also shows 
the KPNO 0.9-m data used by \citet{lopez2014}. Note that the true inner structure is lost to both 
Gemini and KPNO 0.9-m under poor seeing conditions. 

The Gemini data presented in Fig.~\ref{abell85sb} allow us to quantify the impact of seeing. 
The surface brightness profile with excellent seeing and the Gemini profile with poor seeing 
have a small, but measurable, difference of 0.01 mag/arcsec at $r=4.5$ kpc. This difference 
increases to 0.02 mag/arcsec at $r=2.0$ kpc, and to 0.04 mag/arcsec at $r=1.0$ kpc. As shown in 
Fig.~\ref{abell85sb}, the two Gemini profiles begin to diverge at  about $\sim$1.5 kpc 
from the nucleus of the galaxy. 

The fact that the profile obtained with bad seeing fails to accurately record the true 
profile of the galaxy at such large radii is crucial. \citet{lopez2014} claims that the 
Abell 85 BCG has a cusp radius of $r_{\gamma}=4.57 \pm 0.06$ kpc. This value is within the 
range where the surface brightness profile is affected by seeing effects. It is precisely 
this cusp radius that is used to infer the presence of a supermassive black hole.

\section{Final remarks}

A new VLA map shows bipolar AGN jets aligned along the north-south direction in the Abell 85 BCG. 
X-ray and radio images show no evidence of a binary black hole in the core of this galaxy. 
The optical spectra shown here demonstrate that the Abell 85 BCG and SDSS J004150.75-091824.3 are 
two entirely different objects with no relation whatsoever, other than their close projected 
proximity in the sky.

Studies searching for light deficits associated with SMBHs in the inner regions 
of surface brightness profiles must use high resolution data. Poor seeing distorts the central 
region of any surface brightness profile creating the mirage hallmarks of a SMBH. When possible, 
the ideal surface brightness profile should be created by combining HST data in the core 
and ground based data on the outskirts of any given galaxy.

\acknowledgments
We thank Yuto Ichinohe of the Japanese Space Agency (JAXA) for providing the 
exquisite X-ray data. We thank the referee for providing a prompt and constructive report.

This research has made use of the VizieR catalogue access tool, CDS, Strasbourg, France.
Based on data from the GTC Public Archive at CAB (INTA-CSIC).
Based  on observations  obtained at  the Gemini  Observatory  which is 
operated by AURA under a cooperative  agreement with  the NSF  on 
behalf  of the Gemini partnership. 
The National Radio Astronomy Observatory is a facility of the National 
Science Foundation operated under cooperative agreement by Associated 
Universities, Inc.

\software: Astropy \citep{astropy2013}, Matplotlib \citep{hunter2007}, Numpy \citep{vanderwalt2011}.
Common Astronomy Software Applications (CASA) \citep{mcmullin2016}.

\facilities  Gemini, Chandra, VLA, Grantecan.



\end{document}